\documentclass[aps,prd,floatfix,preprint,nofootinbib]{revtex4-1}
\usepackage{graphicx}
\usepackage{epic}
\usepackage{eepic}
\usepackage{latexsym}
\usepackage{amssymb,amsmath}

\newcommand{\eq}[1]{(\ref{#1})}
\newcommand{\be}{\begin{equation}}
\newcommand{\ee}{\end{equation}}
\newcommand{\bea}{\begin{eqnarray}}
\newcommand{\eea}{\end{eqnarray}}

\newcommand{\hs}[1]{\hspace{#1 mm}}

\newcommand{\lf}{\left<}
\newcommand{\rg}{\right>}

\def\d{\delta}

\def\e{\epsilon}

\def\f{\phi}
\def\fr{\frac}
\def\F{\Phi}
\def\vf{\varphi}

\def\s{\sigma}

\def\z{\zeta}

\def\del{\partial}

\let\bm=\bibitem
\def\nn{\nonumber}

\begin{document}

\title{Infrared Logarithms in Minisuperspace Inflation} 

\author{Ali Kaya}
\email[]{ali.kaya@boun.edu.tr}
\affiliation{Bo\~{g}azi\c{c}i University, Department of Physics, 34342, Bebek, \.{I}stanbul, Turkey}

\date{\today}

\begin{abstract}
We examine the emergence of the infrared logarithms in the cosmological perturbation theory applied to the minisuperspace scalar slow-roll inflation. Not surprisingly, in the single scalar field model the curvature perturbation $\z$ is conserved and no $\ln a_B$ behavior appears, where $a_B(t)$ is the background scale factor of the universe. On the other hand, in the presence of a spectator scalar the $n$'th order perturbation theory gives an $(\ln a_B)^n$ correction to $\z$. However, a nonperturbative estimate shows that $\z$ actually becomes the sum of  a constant and a mildly evolving $\ln a_B$ pieces. 
\end{abstract}

\maketitle

\section{Introduction}

Strong infrared behavior is a characteristic feature of the inflationary physics \cite{w1} (for reviews see e.g. \cite{r1,r2}). The modes are continuously pushed from subhorizon to superhorizon regimes, which enlarges short distance quantum effects on cosmologically interesting scales. Most of the time, the loop corrections contain infrared infinities that must properly be handled by viable physical reasoning. The so called infrared logarithms\footnote{In the presence of the entropy perturbations the infrared loop divergences are power law rather than logarithmic, see \cite{pl}.} show up in loops as a reminiscent of the peculiar infrared behavior  (see e.g. \cite{il1,il2,il3,il4,il5,il6,il7,il8,il9,il10,il11}).  

In this paper, we apply  cosmological perturbation theory to the standard scalar slow-roll inflationary model in the minisuperspace approximation. The minisuperspace theory is expected to capture the dynamics of the zero modes of the full theory. Besides, it is free from loop infinities and renormalization issues, which are intricate in the presence of gravity, and it still contains special features related to the gauge invariance and nonlinearities. Therefore the results obtained in the minisuperspace approximation must shed light on some crucial questions in the full theory and our aim here is to examine the appearance of the infrared logarithms.  

We first consider the single scalar field model and obtain the {\it complete} gauge fixed action for the curvature perturbation $\z$. The corresponding Hamiltonian can be expanded in powers of the momentum conjugate to $\z$, which becomes an expansion in the inverse powers of the background scale factor of the universe $a_B(t)$. Asymptotically at late times, the quadratic momentum term, which is still nonlinear in $\z$, dominates the dynamics. In this case, $\z$ is conserved and no $\ln a_B$ behavior appears, which is consistent with \cite{zc1,zc2}. We then add a self-interacting spectator scalar to the system. This time there arises a specific asymptotically dominant interaction term, which yields an $(\ln a_B)^n$ correction to $\z$ in the $n$'th order perturbation theory. The emergence of this infrared logarithm is similar to what has been observed in the field theory calculations. For models having large number of e-folds such a correction may invalidate the perturbation theory. On the other hand, in the minisuperspace theory a nonperturbative argument shows that asymptotically $\z$ has actually a slowly evolving $\ln a_B$ correction to the constant mode, which indicates other infrared logarithms involving higher powers of $\ln a_B$ might be the artifacts of the perturbation theory.   

\section{Single Scalar Slow-roll Inflation}

We start from the following minisuperspace action:
\be\label{1}
S=L^3\int dt\, a^3 \, \left\{\fr{1}{N} \left[-6\fr{\dot{a}^2}{a^2}+\fr12 \dot{\F}^2\right]-NV(\F)\right\},
\ee
where the dot denotes the time derivative, $N$ is the lapse function, $a(t)$ is the scale factor of the universe, $\F$ is the inflaton and $V(\F)$ is the inflaton potential (we set the reduced Planck mass $M_p=1$. The proper $M_p$ factors can easily be reinstated by dimensional analysis as we will do below). The minisuperspace action \eq{1} can be obtained from the usual Einstein-Hilbert action by setting $N^i=0$, $h_{ij}=a^2\d_{ij}$, where $N$, $N^i$ and $h_{ij}$ refer to the standard ADM decomposition of the metric, and by assuming that the variables $N$, $a$ and $\F$ depend only on time. The parameter $L$ denotes the size of the comoving spatial coordinates and the factor $L^3$ in \eq{1} arises from their integration reducing the field theory to a quantum mechanical system. 

The action \eq{1} is invariant under a local time transformation with the parameter $k^0$:
\be\label{2}
\d N=k^0\dot{N}+N\dot{k}^0,\hs{10}\d\F=k^0\dot{\F},\hs{10}\d a=k^0\dot{a}.
\ee
To fix the gauge invariance one may define the background field variables $a_B$ and $\F_B$ obeying
\be\label{3}
6H_B^2=\fr12\dot{\F}_B^2+V(\F_B),\hs{10}\dot{H}_B^2=-\fr14\dot{\F}_B^2,
\ee
where $H_B=\dot{a}_B/a_B$. Next, one may introduce the fluctuation fields $\z$ and $\f$ as
\be\label{4}
a=a_Be^{\z},\hs{10}\F=\F_B+\f,
\ee
and impose the gauge $\f=0$. After algebraically solving the lapse $N$  from its own equation of motion one may obtain 
\be\label{5}
S=-2L^3\int dt\,a_B^3V_B\,e^{3\z}\,\left[1+\fr{12H_B}{V_B}\dot{\z}+\fr{6}{V_B}\dot{\z}^2\right]^{1/2},
\ee
where $V_B=V(\F_B)$. We take the background \eq{3} to be a slow-roll inflationary solution with $\dot{\F}_B<0$. 

Assuming that the physical size of the pre-inflationary patch is determined by the initial Hubble parameter $H_i$, one has
\be\label{6}
a_B(t_i)L=\fr{1}{H_i},
\ee
where $t_i$ is the initial time of the inflation. Normalizing the scale factor as 
\be\label{7}
a_B(t_i)=\fr{1}{H_i}
\ee
corresponds to setting $L=1$, which eliminates the unphysical comoving scale from the equations. 

It is instructive to repeat  the gauge fixing procedure in the Hamiltonian formulation where the minisuperspace action \eq{1} can be written as 
\bea
&&S=\int dt\left[\hat{P}_\z \dot{\hat{\z}}+P_\F\dot{\F}-NH\right],\nn\\
&&H=-\fr{1}{24}e^{-3\hat{\z}}\hat{P}_\z^2+\fr12e^{-3\hat{\z}}P_\F^2+e^{3\hat{\z}}V(\F)\label{8},
\eea
and $\hat{\z}=\ln a$. Expanding the variables around their background values
\bea
&&\hat{\z}=\ln a_B+\z,\hs{10}\F=\F_B+\f,\hs{10}N=1+n,\nn\\
&&\hat{P}_\z=-12a_B^3H_B+P_\z,\hs{10}P_\F=a_B^3\dot{\F}_B+P_\f,\label{9}
\eea 
the action becomes
\be
S=\int dt \left[P_\z\dot{\z}+P_\f\dot{\f}-H_F-nC\right],\label{10}
\ee
where $H_F$ is the fluctuation Hamiltonian involving all variables but $n$ and $C$ is the constraint given by
\be\label{11}
C=-\fr{1}{24a_B^3}e^{-3\z}\left(-12a_B^3H_B+P_\z\right)^2+\fr{1}{2a_B^3}e^{-3\z}\left(a_B^3\dot{\F}_B+P_\f\right)^2+a_B^3e^{3\z}V(\F_B+\f).
\ee
After imposing the gauge $\f=0$, one may solve\footnote{In solving $P_\f$ one should keep in mind that $a_B^3\dot{\F}_B+P_\f<0$ since we are making an expansion around a background solution with $\dot{\F}_B<0$.} the constraint $C=0$ for $P_\f$, which would give the reduced action for $\z$ and $P_\z$ 
\be\label{12}
S=\int\, dt\, P_\z\dot{\z}-\dot{\F}_B\left[\fr{1}{12}\left(P_\z-12a_B^3H_B\right)^2-2a_B^6V_Be^{6\z}\right]^{1/2}+H_BP_\z+6a_B^3V_B\z.
\ee
In the phase space path integral quantization, this procedure corresponds to the Faddeev-Popov gauge fixing
\be\label{13}
\d(\f)\d(C) \det\left\{\f,C\right\}=\d(\f)\d(C)\fr{\del C}{\del\f}=\d(\f)\d(P_\f-P_\f^*),
\ee
where $P_\f^*$ is the solution of $C=0$. One can check that the two actions \eq{5} and \eq{12} are related by the Legendre transformation exchanging the Lagrangian and the Hamiltonian. 

One may see that a constant $\z$ solves the equations of motion that follows from \eq{5} provided that the background equations \eq{3} are satisfied. At first, this is not obvious from \eq{5} since it contains a pure $\z$ term with no time derivatives when the square root in \eq{5} is expanded. In any case, it is possible to add \eq{5} a total derivative term so that
\bea
S&=&-2\int dt\,a_B^3V_B\,e^{3\z}\,\left[1+\fr{12H_B}{V_B}\dot{\z}+\fr{6}{V_B}\dot{\z}^2\right]^{1/2}+2\int dt \,a_B^3\,e^{3\z}\left[V_B+6H_B\dot{\z}\right],\label{14}\\
&=&\int dt\, a_B^3\,e^{3\z}\left[\fr{3\dot{\F}_B^2}{V_B}\dot{\z}^2-\fr{18H_B\dot{\F}_B^2}{V_B^2}\dot{\z}^3+. . . \right].\label{15}
\eea
It is now clear from \eq{15} that the equation of motion involves only the time derivatives of $\z$ and a constant mode  is a trivial solution. Note that by normalization \eq{7}, the scale factor $a_B$ has mass dimension $-1$. In the Hamiltonian language the extra surface term added in \eq{14} corresponds to a canonical transformation $P_\z\to P_\z+12a_B^3H_B-12 a_B^3H_B e^{3\z}$ as compared to \eq{12}. The Hamiltonian of \eq{14} can be found as 
\bea
H&=&-a_B^3\dot{\F}_B^2e^{3\z}\left[1-\fr{2H_B}{a_B^3\dot{\F}_B^2}e^{-3\z}P_\z+\fr{1}{12a_B^6\dot{\F}_B^2}e^{-6\z}P_\z^2\right]^{1/2}-H_B P_\z+a_B^3\dot{\F}_B^2e^{3\z},\label{16}\\
&=&\fr{V_B}{12a_B^3\dot{\F}^2}e^{-3\z}P_\z^2+\fr{H_BV_B}{12a_B^6\dot{\F}_B^4}e^{-6\z}P_\z^3+. . . \label{17}
\eea
where the dotted terms are suppressed with more powers of the background scale factor. Evidently, the first term in \eq{17} dominates the dynamics at late times in inflation. 

The interaction picture operators are governed by the free Hamiltonian
\be\label{18}
H_0=\fr{V_B}{12a_B^3\dot{\F}_B^2}P_\z^2.
\ee
Their time evolution can be found as
\bea
&&\z_I(t)=\z_i+\fr16\int_{t_i}^t dt'\fr{V_B(t')}{a_B(t')^3\dot{\F}_B(t')^2}\,P_i,\nn\\
&&P_{\z I}=P_i,\label{19}
\eea
where $\z_i$ and $P_i$ are the initial time independent (Schr\"{o}dinger) operators obeying $[\z_i,P_i]=i$. 

An operator in the Heisenberg picture $O_H$ can be related to the corresponding interaction picture operator $O_I$ by
\be
O_H=U_I^\dagger O_I U_I,\label{20}
\ee
where $i\dot{U}_I=H_IU_I$, $U(t_i)=I$ and $H_I$ is the interaction Hamiltonian in the interaction picture. As shown by Weinberg \cite{il1}, \eq{20} can be expanded as
\be\label{21}
O_H(t)=O_I(t)-i\int_{t_i}^t dt'\,[O_I(t),H_I(t')]-\int_{t_i}^t dt''\int_{t''}^t dt'\,[[O_I(t),H_I(t')],H_I(t'')]+. . . 
\ee
where the dotted terms contain more nested commutators of $O_I$ with $H_I$. Eq. \eq{21} can be used as the basis for the in-in perturbation theory. From \eq{17} the interaction Hamiltonian can be determined as
\be\label{22}
H_I=\fr{V_B}{24a_B^3\dot{\F}_B^2}\left\{\left(e^{3\z_I}-1\right),P_{\z I}^2\right\}+. . .
\ee
where we apply symmetric ordering to make $H_I$ Hermitian. 

One may approximate the time integrals during slow-roll inflation by taking (note the normalization \eq{7}) 
\be
a_B\simeq \fr{1}{H_B}e^{H_B(t-t_i)}\label{23}
\ee
and by treating the slowly changing variables $H_B$, $V_B$ and $\dot{\F}_B$ as constants. Using \eq{22} in \eq{21} for $\z$, one finds that at the end of inflation after $N$ e-folds 
\be\label{24}
\z_H=\z_i+\fr{H_B^2}{12M_p^2\e}P_i+. . .+O\left(e^{-3N}\right),
\ee
where the slow-roll parameter is defined as
\be
\e=\fr{3\dot{\F}_B^2}{V_B}\simeq-\fr{\dot{H}_B}{H_B},\label{25}
\ee
and dots denote time independent but nonlinear terms in $\z_i$ and $P_i$ coming from the lower limits of the time integrals in \eq{21} at $t_i$. Consequently, one sees that at late times $\z_H$ exponentially asymptotes to a constant operator and no infrared logarithms appear.

We observe that neglecting all but the first term in \eq{17}, which are exponentially suppressed at late times, gives an explicitly integrable system. Namely, the (classical) equations corresponding to the Hamiltonian
\be
H=\fr{V_B}{12a_B^2\dot{\F}_B^2}e^{-3\z}P_\z^2,\label{26}
\ee
can be integrated to get
\bea
&&\z(t)=\z_i+\fr23 \ln\left[1+P_i e^{-3\z_i}\int_{t_i}^t dt'\fr{V_B}{4a_B^3\dot{\F}_B^2}\right],\nn\\
&&P_\z(t)=P_i+P_i^2e^{-3\z_i}\int_{t_i}^t dt'\fr{V_B}{4a_B^3\dot{\F}_B^2}.\label{27}
\eea
In the quantum theory \eq{27} should be true for Heisenberg operators provided that operator orderings are solved in a suitable way. Eq. \eq{27} shows that the asymptotic change of $\z$ compared to its initial value is determined by the dimensionless parameter $H^2/(M_p^2\e)$. 

Till now in our discussion we have focused on the evolution of the Heisenberg operator $\z_H$. As for the initial state it is natural to take a minimum uncertainty Gaussian wave function $\psi(\z_i)$, which has zero mean $<\z_i>=0$ and the deviations $<\z_i>=\s^2$, $<P_i^2>=1/4\s^2$. Although this choice can be motivated from field theory side, the value of the deviation $\s$ cannot be directly deduced from the field theory, which has a continuous spectrum of wave numbers and the zero mode is not isolated (unless the space is not compact). On the other hand, one must also note that the validity of the perturbation theory actually depends on the initial state. Choosing $\s$ to be extremely small would yield a large momentum that may invalidate the perturbative expansion in $P_\z$, at least at early times during inflation when the exponential suppression is not effective yet. 

In showing the constancy of $\z$ in single scalar inflationary models, the consistency condition in the squeezed limit \cite{c1,c2}, hence the choice of the Bunch-Davies vacuum, plays an important role, see \cite{c3}. In the minisuperspace model, no such property is needed since the time independence of $\z$ becomes an operator statement, i.e. the Heisenberg picture $\z_H$ exponentially approaches to a constant operator as in \eq{24}. We anticipate this should also be the case in field theory since at late times semi-classical approximation becomes excellent \cite{cl} and $\z$ is conserved in the classical theory.\footnote{Indeed, one naturally expects that some form of minisuperspace description of superhorizon modes, which is similar to the one considered here, must be valid at late times. However, such an approximation, if ever exists, is only possible in a suitable gauge that allows a smooth soft limit, which is not the case for the standard $\z$-gauge because the shift $N^i$ is non-local.} Therefore, the constancy of $\z$ in cosmological perturbation theory must hold not just for the Bunch-Davies vacuum but for a wider range of states. 

\section{Adding a Spectator}

We have seen in the previous section that the $\z$-self interactions cannot yield infrared logarithms in the minisuperspace perturbation theory. From \eq{19} and \eq{23} one sees that $[\z_I(t),\z_I(t')]\propto 1/a_B^3$, thus the time integrals in the perturbative series in \eq{21} can produce an infrared logarithm provided that $H_I\propto a_B^3$. To produce such an interaction term one may add a self-interacting {\it massless} spectator scalar $\vf$ which has the potential $V(\vf)$. It is easy to repeat the gauge fixing in the presence of the spectator to get the following gauge fixed action: 
\be
S=-2\int dt\,a_B^3V_B\,e^{3\z}\,\left[1+\fr{V(\vf)}{V_B}\right]^{1/2}\left[1+\fr{12H_B}{V_B}\dot{\z}+\fr{6}{V_B}\dot{\z}^2-\fr{1}{2V_B}\dot{\vf}^2\right]^{1/2}.\label{28}
\ee
By expanding the square roots one may obtain the free Lagrangian and various interactions, where the quadratic spectator action is given by
\be\label{29}
S=\fr12\,\int \,dt\, a_B^3\,\dot{\vf}^2.
\ee
Hence, the interaction picture spectator operators evolve like 
\bea
&&\vf_I=\vf_i+\int_{t_i}^t \fr{dt'}{a_B(t')^3}P_{\vf i},\nn\\
&&P_{\vf I}=P_{\vf i},\label{30}
\eea
where $P_{\vf I}$ is the momentum conjugate to $\vf_I$, and $\vf_i$ and $P_{\vf i}$ are time independent initial operators obeying $[\vf_i,P_{\vf i}]=i$. 

Among the interactions that follow from \eq{28} we focus on the following one
\be
H_I=a_B^3V(\vf_I)\left(e^{3\z_I}-1\right),\label{31}
\ee
which would potentially yield infrared logarithms in the perturbation theory as noted above. Indeed, from the first order correction in \eq{21} one may find
\be\label{32}
\z_H=\z_I-\fr12 \int_{t_i}^t dt'a_B(t')^3\,V(\vf_I(t'))e^{3\z_I(t')}\int_{t'}^t\,dt''\fr{V_B}{a_B^3\dot{\F}_B^2}.
\ee
Using \eq{23} one can get a late time expansion of the interaction picture operators so that at the end of inflation after $N$ e-folds one has 
\bea
&&\z_I=\z_i+\fr{H_B^2}{12M_p^2\e}P_{i}+O\left(e^{-3N}\right),\nn\\
&& \vf_I=\vf_i+\fr{H_B^2}{3}P_{\vf i}+O\left(e^{-3N}\right).\label{33}
\eea
Utilizing this expansion in \eq{32} gives
\be
\z_H(t)=\z_c+\fr{1}{12H_B^2M_p^2\e}\left[1-3\ln\left(\fr{a_B(t)}{a_B(t_i)}\right)\right]\,V(\vf_c)\,e^{3\z_c}+O\left(e^{-3N}\right),\label{34}
\ee
where the constant operators $\z_c$ and $\vf_c$ are defined from \eq{33} by
\bea
&&\z_c=\z_i+\fr{H_B^2}{12M_p^2\e}P_{i},\nn\\
&&\vf_c=\vf_i+\fr{H_B^2}{3}P_{\vf i}.\label{344}
\eea
As it is anticipated, the interaction \eq{31} yields an infrared logarithm in the first order perturbation theory. 

The above calculation hints how one should handle the higher order perturbative corrections. Namely, one should first evaluate the commutators in \eq{21} using
\be
[\z_I(t),\z_I(t')]=\fr{i}{6}\int_t^{t'}dt''\fr{V_B}{a_B^3\dot{\F}_B^2},\hs{10}[\vf_I(t),\vf_I(t')]=i\int_t^{t'}\fr{dt''}{a_B^3}.\label{35}
\ee
One can then apply the late time expansion of the interaction picture operators given in \eq{33} and calculate the time integrals of the leading order terms. Using this strategy one may obtain the following second order correction to $\z_H$:
\be
\left[\fr{V(\vf_c)^2}{16M_p^4H_B^2\e^2}e^{6\z_c}+\fr{V'(\vf_c)^2}{36M_p^2H_B^2\e}\left(e^{6\z_c}-e^{3\z_c}\right)\right]\left[1-2\ln\left(\fr{a_B(t)}{a_B(t_i)}\right)+\fr32 \ln^2\left(\fr{a_B(t)}{a_B(t_i)}\right)\right],\label{36}
\ee
where $V'(\vf)=dV/d\vf$. Note that \eq{36} contains a different type of infrared logarithm, i.e. a log square. 

It is possible to argue that the interaction \eq{31} yields the factor $\ln^n(a_B(t)/a_B(t_i))$ in the $n$'th order perturbation theory. In the $n$'th term of \eq{21} there are $n$ factors of $a_B^3$ coming from the interaction Hamiltonian and there are $n$ factors of $a_B^{-3}$ coming from $n$-commutators. At late times, these cancel each other and the interaction picture operators asymptote to constant operators as in \eq{33}. Hence, to leading order one ends up with an $n$-dimensional time integral of a constant operator giving $(t-t_i)^n=H_B^n\ln^n(a_B(t)/a_B(t_i))$. 

These findings are consistent with what has been observed in the field theory calculations \cite{il1,il2,il3}. In the minisuperspace approximation one can further make a nonperturbative estimate as follows: Using the asymptotic form \eq{33}, the interaction Hamiltonian converges to
\be\label{37}
H_I=a_B^3V(\vf_c)\left(e^{3\vf_c}-1\right)\left\{1+O\left(e^{-3N}\right)\right\}.
\ee
One can check that at late times the commutator $[H_I(t),H_I(t')]$ is suppressed by a huge factor related to the number of e-folds as compared to the product $H_I(t)H_I(t')$. Therefore, to a very good approximation $H_I(t)$ becomes a self-commuting operator of its argument after some time $t_m$ corresponding to, say, 10 e-folds (the fact that $\z$ has a similar property has been used to argue the classicality of the cosmological perturbations \cite{cl}). The unitary interaction picture evolution operator can be decomposed like
\be
U_I(t,t_i)=U_2(t,t_m)U_1(t_m,t_i).\label{38}
\ee
Since $H_I(t)$ can be treated as a self-commuting operator when $t>t_m$, one may approximate
\be
U_2(t,t_m)=Te^{-i\int_{t_m}^t dt'H_I(t')}\simeq e^{-i\int_{t_m}^t dt'H_I(t')}.\label{39}
\ee
Furthermore one has
\be
\z_H=U_I^\dagger\z_IU_I=U_1^\dagger U_2^\dagger\z_I U_2U_1.\label{40}
\ee
To proceed we note 
\be
[\z_I(t),H_I(t')]=3[\z_I(t),\z_I(t')]H_I(t'),\label{41}
\ee
thus using \eq{39} one may find
\bea
[\z_I(t),U_2]&&\simeq\fr12 \int_{t_m}^tdt'a_B(t')^3V(\vf(t'))e^{3\z_I(t')}\int_t^{t'}dt''\fr{V_B(t'')}{a_B(t'')^3\dot{\F}_B(t'')^2}\,U_2\nn\\
&&\simeq \fr{1}{12H_B^2M_p^2\e}\left[1-3\ln\left(\fr{a_B(t)}{a_B(t_i)}\right)\right]V(\vf_c)e^{3\vf_c}\,U_2.\label{42}
\eea
So \eq{40} becomes
\be
\z_H\simeq U_1^\dagger\z_cU_1+\fr{1}{12H_B^2M_p^2\e}\left[1-3\ln\left(\fr{a_B(t)}{a_B(t_i)}\right)\right]U_1^\dagger V(\vf_c)e^{3\vf_c}U_1.\label{43}
\ee
The unitary operator $U_1$ only mixes the operators up to time $t_m$ and its action merely produces constant operators since these do not depend on the final time. As a result, in \eq{43} we are able to extract the leading order time dependence of $\z$, which is a single infrared logarithm of the form $\ln a_B$, and other corrections are exponentially suppressed. On dimensional grounds one may estimate that (in expectation values) $V(\vf_c)\propto H_B^4$, therefore the infrared logarithm correction is suppressed by the factor $H^2/(M_p^2\e)$, which is generically small in realistic models. 

The above nonperturbative argument shows that the $(\ln a_B)^n$ behavior for $n>1$ that arises in the $n$'th order perturbation theory may be an artifact of that approximation. For $t<t_m$ the sum of all these terms must yield the time evolution dictated by the unitary operator $U_1$, which we manage to bypass in our analysis. The minisuperspace approximation clearly shows the emergence of a single infrared logarithm related to the asymptotic form of the interaction picture operators. Specifically, after some time during inflation an interaction picture operator becomes almost self-commutative when evaluated at different time arguments. In that case, in relating the Heisenberg and the interaction picture operators by a unitary transformation, it is enough to carry out a single commutator that cancels $a_B^3$ factor in the interaction Hamiltonian.  The time integral of the leading order constant operators then gives a single infrared logarithm as in \eq{43}.  

Finally, it is instructive to consider a {\it massive} spectator to see how the emergence of infrared logarithms depends on the mass parameter. In that case the quadratic spectator action \eq{29} becomes 
\be\label{29m}
S=\fr12\,\int \,dt\, a_B^3\,\left[\dot{\vf}^2-m^2\vf^2\right]
\ee
and the interaction picture operators can be solved as 
\bea
&&\vf_I=f_1(t)\vf_i+f_2(t)P_{\vf i},\nn\\
&&P_{\vf I}=a_B^3\dot{f}_1(t)\,\vf_i+ a_B^3\dot{f}_2(t)\,P_{\vf i},\label{30m}
\eea
where $f_1$ and $f_2$ are the two solutions of 
\be\label{de}
\ddot{f}+3H_B\,\dot{f}+m^2\,f=0
\ee
determined by the initial conditions
\bea
&&f_1(t_i)=1,\hs{5}\dot{f}_1(t_i)=0,\nn\\
&&f_2(t_i)=0,\hs{5}\dot{f}_2(t_i)=a_B(t_i)^{-3}.
\eea
Let us now examine the first order correction which is still given by \eq{32}. For the massless spectator when $m=0$, one has $f_1=1$ and $f_2=[a_B(t_i)^{-3}-a_B(t)^{-3}]/(3H_B)$, where we assume \eq{23}. Here, $f_1$ is already a constant and $f_2$ exponentially approaches to a constant, consequently one has \eq{33} which then gives the infrared logarithm \eq{34}. However, when $m\not=0$ the two solutions of \eq{de} for constant $H_B$ are given by
\be\label{sol}
\exp\left[n_\pm H_B(t-t_i)\right],\hs{5}n_\pm=-\fr32\pm\sqrt{\fr94-\fr{m^2}{H_B^2}},
\ee
which are decreasing functions of time. Accordingly, both $f_1$ and $f_2$ turn out to decay either as $\exp(-m^2(t-t_i)/3H_B)$ for $m\ll H_B$ or as $a_B^{-3/2}$ for $m>3H_B/2$. Using \eq{30m} in \eq{32}, one may then see that the integrand in \eq{32} does {\it not} approach to a time independent operator yielding the infrared logarithm as in the case of a massless spectator, but instead it becomes an exponentially decreasing function of $t'$ whose integral gives a smaller correction for larger mass. 

\section{Conclusions}

In this paper we investigate the appearance of the infrared logarithms in the cosmological perturbation theory by studying the scalar slow-roll inflationary model in the minisuperspace approximation, which simplifies the field theoretical system involving gravity to a quantum mechanical one. The minisuperspace  theory is still highly nontrivial because of the nonlinearities and the local gauge invariance related to the time reparametrizations. We obtain the complete gauge fixed action for the curvature perturbation $\z$, both in the single scalar case and when a self-interacting spectator is added. The full action can be expanded around the inflationary background yielding an infinite number of interaction terms. 

In our analysis we focus on the time evolution of the Heisenberg operators, which can be calculated using in-in perturbation theory. Thus, our findings are state independent provided that the expectation values do not break down the series expansion. We verify that in the single scalar case no infrared logarithms appear and $\z$  exponentially asymptotes to a constant operator. In the presence of a spectator we find that the $n$'th order perturbation theory gives an infrared logarithm of the form $(\ln a_B)^n$. Note that supposing the existence of a spectator is not unnatural for inflation; in any model where Higgs is not the inflaton, it actually becomes a self-interacting spectator scalar.

In the minisuperspace approximation it is possible to examine the time evolution of the Heisenberg operators nonperturbatively. Following some time after the beginning of inflation, the interaction picture operators including the interaction Hamiltonian become nearly self-commuting at different times. This allows one to extract the leading order time evolution of $\z$ where all other corrections are exponentially suppressed. In the presence of a spectator, this leading order correction turns out to be a single infrared logarithm $\ln a_B$. It would be interesting to generalize this argument to field theory to understand the structure of the infrared logarithms in the cosmological perturbation theory. 

One usually attributes the emergence of infrared logarithms to loop effects. This is natural since in field theory calculations they normally appear in loop corrections. However, loops do not exist in the minisuperspace approach yet we still encounter infrared logarithms implied by the Heisenberg picture equations of motion. This indicates that neither the loops nor the modes running in them may not be the primary reason for the existence of infrared logarithms. Indeed, consider as an example the three point function $\lf \f(k_1)\f(k_2)\f(k_3)\rg$ of a self-interacting {\it massless} test scalar field. It is not difficult to see that this correlation function is time dependent {\it at tree level} because of a cubic interaction term $H_I=g\, a^3 \int d^3 x\,\f^3$, even at late times when $k_1$, $k_2$ and $k_3$ become superhorizon. Choosing the vacuum and correspondingly the mode functions determine the precise form of this time dependence, e.g. for the Bunch-Davies vacuum one gets an infrared logarithm. The crucial point is that the cubic interaction induces nontrivial superhorizon evolution which is {\it not} exponentially suppressed. In single field inflation, one may see that $\z$ self interactions cannot produce such effects mainly because of the shift symmetry (as shown above, pure $\z$ interactions containing no derivatives, which are potentially dangerous, actually disappear after integration by parts). Nevertheless, in the presence of a spectator field there are interactions which yield non-negligible superhorizon evolution both in classical and quantum theories. We expect these conclusions to hold irrespective of the gauge conditions or possible explicit non-localities present in the action. 

Hence, the emergence of infrared logarithms or some other form of nontrivial superhorizon motion has a dynamical origin related to suitable interactions. Remember that in the minisuperspace model, solving the unitary evolution nonperturbatively gives only a single infrared logarithm as opposed to perturbation theory and this shows the importance of determining dynamics correctly. On the other hand, the initial state chosen is also crucial in fixing the exact form of the superhorizon time dependence. Of course, loops also arise from such interactions sandwiched between states. As discussed in \cite{proj}, some corrections related to superhorizon modes are projection effects that define the mapping between physical scales of inflation and post-inflationary universe, and these disappear in observable quantities. One way of understanding projection effects is to keep in mind that the decomposition of the metric into  background and  fluctuation parts is introduced for computational convenience. Strictly, the {\it full} metric must be used in relating the comoving and physical scales. In the minisuperspace theory any constant piece of $\z$ will disappear as a projection artifact but a time dependent part represents a real physical effect, modifying for instance the Hubble expansion rate.

\begin{acknowledgments}
This work, which has been done at C.\.{I}.K. Ferizli, Sakarya, Turkey without any possibility of using references, is dedicated to my friends at rooms C-1 and E-10 who made my stay bearable at hell for 440 days between 7.10.2016 and 20.12.2017. I am also indebted to the colleagues who show support in these difficult times. 
\end{acknowledgments}

\end{document}